\documentclass[aps,pra,showpacs,superscriptaddress,preprint]{revtex4}

\usepackage{graphicx}

\catcode`ð=\active

 \defð{\u{g}}

 \catcode`Ð=\active

 \defÐ{\u{G}}

 \catcode`Ý=\active

 \defÝ{\. I}

 \catcode`ö=\active

 \defö{\"{o}}

 \catcode`Ö=\active

 \defÖ{\"O}

 \catcode`ü=\active

 \defü{\"{u}}

 \catcode`Ü=\active

 \defÜ{\"{U}}

 \catcode`Þ=\active

 \defÞ{\c{S}}

 \catcode`þ=\active

 \defþ{\c{s}}

 \catcode`ý=\active

 \defý{{\i}}

 \catcode`ç=\active

 \defç{\d{c}}

 \catcode`Ç=\active

 \defÇ{\d{C}}

\begin{document}

\title{Effective Mass Dirac-Morse Problem with any $\kappa$-value}

\author{\small Altuð Arda}

\email[E-mail: ]{arda@hacettepe.edu.tr}\affiliation{Department of
Physics Education, Hacettepe University, 06800, Ankara,Turkey}
\author{\small Ramazan Sever}
\email[E-mail: ]{sever@metu.edu.tr}\affiliation{Department of
Physics, Middle East Technical  University, 06531, Ankara,Turkey}
\author{\small Cevdet Tezcan}
\email[E-mail: ]{ctezcan@baskent.edu.tr}\affiliation{Faculty of
Engineering, Baþkent University, Baglýca Campus, Ankara,Turkey}
\author{\small H\"{u}seyin Akçay}
\email[E-mail: ]{akcay@baskent.edu.tr}\affiliation{Faculty of
Engineering, Baþkent University, Baglýca Campus, Ankara,Turkey}

\date{\today}

\begin{abstract}

The Dirac-Morse problem are investigated within the framework of
an approximation to the term proportional to $1/r^2$ in the view
of the position-dependent mass formalism. The energy eigenvalues
and corresponding wave functions are obtained by using the
parametric generalization of the Nikiforov-Uvarov method for any
$\kappa$-value. It is also studied the approximate energy
eigenvalues, and corresponding wave functions
in the case of the constant-mass for pseudospin, and spin cases, respectively.
\end{abstract}


\maketitle

\newpage

The investigation of the solutions for quantum mechanical systems
having certain potentials in the case of position-dependent mass
(PDM) [1, 2] has been received great attentions. Many authors have
studied the solutions of different potentials for
spatially-dependent mass, such as hypergeometric type potentials
[3], Coulomb potential [4], $PT$-symmetric kink-like, and inversely
linear plus linear potentials [5]. It is well known that the theory
based on the effective-mass Schr\"{o}dinger equation is a useful
ground for investigation of some physical systems, such as
semiconductor heterostructures [6], the impurities in crystals
[7-9], and electric properties of quantum wells, and quantum dots
[10]. In the present work, we tend to solve the Dirac-Morse problem
within the PDM formalism.

The pseudospin symmetry is an interesting result appearing in Dirac
equation of a particle moving in an external scalar, and vector
potentials in the case of it when the sum of the potentials is
nearly zero. It was observed that the single particle states have a
quasidegeneracy labeled with the quantum numbers $\tilde{\ell}$, and
$\tilde{s}$, which are called the pseudo-orbital angular momentum,
and pseudospin angular momentum quantum numbers, respectively
[11-16]. The concept of pseudospin symmetry has received great
attentions in nuclear theory because of being a ground to
investigate deformation, and superdeformation in nuclei [17, 18],
and to build an effective shell-model coupling scheme [19, 20]. The
symmetry appears in that case, when the magnitude of scalar
potential is nearly equal to the magnitude of vector potential with
opposite sign [14, 21-25] and the Dirac equation has the pseudospin
symmetry, when the sum of the vector, and scalar potentials is a
constant, i.e., $\Sigma(r)=V_v(r)+V_s(r)=const.$ or
$d\Sigma(r)/dr=0$ [16]. The spin symmetry is another important
symmetry occurring in Dirac theory in the presence of external
scalar, and vector potentials. The spin symmetry appears in the
Dirac equation, when the difference of scalar, and vector potentials
is a constant, i.e., $\Delta(r)=V_{v}(r)-V_{s}(r)=const.$ [14, 16].

Recently, the pseudospin and/or spin symmetry have been studied by
many authors for some potentials, such as Morse potential [26-28],
Woods-Saxon potential [29], Coulomb [30], and harmonic potentials
[31-33], Eckart potential [34-36], P\"{o}schl-Teller potential[37,
38], Hulth\'{e}n potential [39], and Kratzer potential [40]. In
Ref. [41], the bound-state solutions of Dirac equation are studied
for generalized Hulth\'{e}n potential with spin-orbit quantum
number $\kappa$ in the position-dependent mass background. In this
letter, we tend to show that the new scheme of the
Nikiforov-Uvarov (NU) method could be used to find the energy
spectra, and the corresponding eigenspinors within the framework
of an approximation to the term proportional to $1/r^2$ for
arbitray spin-orbit quantum number $\kappa$, i.e. $\kappa\neq 0$,
when the mass depends on position. The NU method is a powerful
tool to solve of a second order differential equation by turning
it into a hypergeometric type equation [42].

Dirac equation for a spin-$\frac{1}{2}$ particle with mass $m$
moving in scalar $V_s(r)$, and vector potential $V_v(r)$ can be
written as (in $\hbar=c=1$ unit)
\begin{eqnarray}
[\alpha\,.\,\textbf{P}+\beta(m+V_s(r))]\,\Psi_{n\kappa}(r)=[E-V_v(r)]\,\Psi_{n\kappa}(r)\,.
\end{eqnarray}
where $E$ is the relativistic energy of the particle, $\textbf{P}$
is three-momentum, $\alpha$ and $\beta$ are $4 \times 4$ Dirac
matrices, which have the forms of $\alpha=\Bigg(\begin{array}{cc}
 0 & \sigma \\
\sigma & 0
\end{array}\Bigg)$ and $\beta=\Bigg(\begin{array}{cc}
0 & I \\
-I & 0
\end{array}\Bigg)$, respectively, [43].
Here, $\sigma$ is a three-vector whose components are Pauli
matrices and $I$ denotes the $2 \times 2$ unit matrix.
$\textbf{J}$ denotes the total angular momentum , and
$\hat{K}=-\beta(\sigma.\textbf{L}+1)$ corresponds to the
spin-orbit operator of the Dirac particle in a spherically
symmetric potential, where $\textbf{L}$ is the orbital angular
momentum operator of the particle. The eigenvalues of the
spin-orbit operator $\hat{K}$ are given as $\kappa=\pm(j+1/2)$,
where $\kappa=-(j+1/2)<0$ correspond to the aligned spin
$j=\ell+1/2$, and $\kappa=(j+1/2)>0$ correspond to the unaligned
spin $j=\ell-1/2$. The total angular momentum quantum number of
the particle is described as $j=\tilde{\ell}+\tilde{s}$\,,where
$\tilde{\ell}=\ell+1$ is the pseudo-orbital angular momentum
quantum number, and $\tilde{s}=1/2$ is the pseudospin angular
momentum quantum number. For a given $\kappa=\pm1, \pm2, \ldots$,
the relation between the spin-orbit quantum number $\kappa$\,, and
"two" orbital angular momentum quantum numbers are given by
$\kappa(\kappa+1)=\ell(\ell+1)$, and
$\kappa(\kappa-1)=\tilde{\ell}(\tilde{\ell}+1)$.

The Dirac spinor in spherically symmetric potential can be written
in terms of upper and lower components as
\begin{eqnarray}
\Psi_{n \kappa}(r)=\,\frac{1}{r}\,\Bigg(\begin{array}{c} \,\chi_{n
\kappa}\,(r)Y_{jm}^{\ell}(\theta,\phi) \\
i\phi_{n \kappa}\,(r)Y_{jm}^{\tilde{\ell}}(\theta,\phi)
\end{array}\Bigg)\,,
\end{eqnarray}
where $Y_{jm}^{\ell}(\theta,\phi)$, and
$Y_{jm}^{\tilde{\ell}}(\theta,\phi)$ are the spherical harmonics,
and $\chi_{n \kappa}\,(r)/r$, and $\phi_{n \kappa}\,(r)/r$ are
radial part of the upper and lower components. Substituting Eq.
(2) into Eq. (1) enable us to write the Dirac equation as a set of
two couple differential equations in terms of $\chi_{n
\kappa}\,(r)$ and $\phi_{n \kappa}\,(r)$. By eliminating $\chi_{n
\kappa}\,(r)$ or $\phi_{n \kappa}\,(r)$ in these coupled
equations, we obtain
\begin{eqnarray}
\Big\{\,\frac{d^2}{dr^2}-\,\frac{\kappa(\kappa+1)}{r^2}\,
+\,\frac{1}{M_{\Delta}(r)}\Big(\frac{dm(r)}{dr}
-\frac{d\Delta(r)}{dr}\Big)\,(\frac{d}{dr}\,+\,\frac{\kappa}{r})\Big\}\chi_{n\kappa}(r)=
M_{\Delta}(r)M_{\Sigma}(r)\chi_{n\kappa}(r)\,,
\end{eqnarray}
\begin{eqnarray}
\Big\{\,\frac{d^2}{dr^2}-\,\frac{\kappa(\kappa-1)}{r^2}\,
-\,\frac{1}{M_{\Sigma}(r)}\Big(\frac{dm(r)}{dr}+
\frac{d\Sigma(r)}{dr}\Big)\,(\frac{d}{dr}\,-\,\frac{\kappa}{r})\Big\}\phi_{n\kappa}(r)=
M_{\Delta}(r)M_{\Sigma}(r)\phi_{n\kappa}(r)\,,
\end{eqnarray}
where $M_{\Delta}(r)=m+E_{n\kappa}-\Delta(r)$\,,
$M_{\Sigma}(r)=m-E_{n\kappa}+\Sigma(r)$, and
$\Delta(r)=V_{v}\,(r)-V_s\,(r)$, $\Sigma(r)=V_{v}\,(r)+V_s\,(r)$.

In the NU-method, the Schr\"{o}dinger equation is transformed by
using an appropriate coordinate transformation
\begin{eqnarray}
\sigma^{2}(s)\Psi''(s)+\sigma(s)\tilde{\tau}(s)
\Psi'(s)+\tilde{\sigma}(s)\Psi(s)=0\,,
\end{eqnarray}
where $\sigma(s)$, $\tilde{\sigma}(s)$ are polynomials, at most
second degree, and $\tilde{\tau}(s)$ is a first degree polynomial.
The polynomial $\pi(s)$, and the parameter $k$ are required in the
method
\begin{eqnarray}
\pi(s)=\frac{1}{2}\,[\sigma^{\prime}(s)-\tilde{\tau}(s)]\pm
\sqrt{\frac{1}{4}\,[\sigma^{\prime}(s)-\tilde{\tau}(s)]^2-
\tilde{\sigma}(s)+k\sigma(s)},
\end{eqnarray}
\begin{eqnarray}
\lambda=k+\pi^{\prime}(s ),
\end{eqnarray}
where $\lambda$ is a constant. The function under the square root
in the polynomial in $\pi(s)$ in Eq. (6) must be square of a
polynomial in order that $\pi(s)$ be a first degree polynomial.
Replacing $k$ into Eq. (6), we define
\begin{eqnarray}
\tau(s)=\tilde{\tau}(s)+2\pi(s).
\end{eqnarray}
where the derivative  of $\tau(s)$ should be negative [42]. Eq.
(5) has a particular solution with degree $n$, if $\lambda$ in Eq.
(7) satisfies
\begin{eqnarray}
\lambda=\lambda_{n}=-n\tau^{\prime}-\frac{\left[n(n-1)\sigma^{\prime\prime}\right]}{2},
\quad n=0,1,2,\ldots
\end{eqnarray}
To obtain the solution of Eq. (5) it is assumed that the solution
is a product of two independent parts as $\Psi(s)=\phi(s)~y(s)$,
where $y(s)$ can be written as
\begin{eqnarray}
y_{n}(s)\sim \frac{1}{\rho(s)}\frac{d^{n}}{ds^{n}}
\left[\sigma^{n}(s)~\rho(s)\right],
\end{eqnarray}
where the function $\rho(s)$ is the weight function, and should
satisfy the condition
\begin{eqnarray}
\left[\sigma(s)~\rho(s)\right]'=\tau(s)~\rho(s)\,,
\end{eqnarray}
and the other factor is defined as
\begin{eqnarray}
\frac{1}{\phi(s)}\frac{d\phi(s)}{ds}=\frac{\pi(s)}{\sigma(s)}.
\end{eqnarray}
In order to clarify the parametric generalization of the NU
method, let us take the following general form of a
Schr\"{o}dinger-like equation written for any potential,
\begin{eqnarray}
\left\{\frac{d^{2}}{ds^{2}}+\frac{\alpha_{1}-\alpha_{2}s}{s(1-\alpha_{3}s)}
\frac{d}{ds}+\frac{-\xi_{1}s^{2}+\xi_{2}s-\xi_{3}}{[s(1-\alpha_{3}s)]^{2}}\right\}\Psi(s)=0.
\end{eqnarray}
When Eq. (13) is compared with Eq. (5), we obtain
\begin{eqnarray}
\tilde{\tau}(s)=\alpha_{1}-\alpha_{2}s\,\,\,;\,\,\sigma(s)=s(1-\alpha_{3}s)\,\,\,;\,\,
\tilde{\sigma}(s)=-\xi_{1}s^{2}+\xi_{2}s-\xi_{3}\,.
\end{eqnarray}
Substituting these into Eq. (6)
\begin{eqnarray}
\pi(s)=\alpha_{4}+\alpha_{5}s\pm\sqrt{(\alpha_{6}-k\alpha_{3})s^{2}+(\alpha_{7}+k)s+\alpha_{8}}\,,
\end{eqnarray}
where the parameter set are
\begin{eqnarray}
\begin{array}{lll}
\alpha_{4}=\frac{1}{2}\,(1-\alpha_{1})\,, &
\alpha_{5}=\frac{1}{2}\,(\alpha_{2}-2\alpha_{3})\,,
& \alpha_{6}=\alpha_{5}^{2}+\xi_{1} \\
\alpha_{7}=2\alpha_{4}\alpha_{5}-\xi_{2}\,, &
\alpha_{8}=\alpha_{4}^{2}+\xi_{3}\,. &
\end{array}
\end{eqnarray}
In NU-method, the function under the square root in Eq. (15) must
be the square of a polynomial [42], which gives the following
roots of the parameter $k$
\begin{eqnarray}
k_{1,2}=-(\alpha_{7}+2\alpha_{3}\alpha_{8})\pm2\sqrt{\alpha_{8}\alpha_{9}}\,,
\end{eqnarray}
where
$\alpha_{9}=\alpha_{3}\alpha_{7}+\alpha_{3}^{2}\alpha_{8}+\alpha_{6}$\,.
We obtain the polynomials $\pi(s)$ and $\tau(s)$ for
$k=-(\alpha_{7}+2\alpha_{3}\alpha_{8})-2\sqrt{\alpha_{8}\alpha_{9}}$,
respectively
\begin{eqnarray}
\pi(s)=\alpha_{4}+\alpha_{5}s-\left[(\sqrt{\alpha_{9}}+\alpha_{3}\sqrt{\alpha_{8}}\,)s-\sqrt{\alpha_{8}}\,\right]\,,
\end{eqnarray}
\begin{eqnarray}
\tau(s)=\alpha_{1}+2\alpha_{4}-(\alpha_{2}-2\alpha_{5})s-2\left[(\sqrt{\alpha_{9}}
+\alpha_{3}\sqrt{\alpha_{8}}\,)s-\sqrt{\alpha_{8}}\,\right].
\end{eqnarray}
Thus, we impose the following for satisfying the condition that
the derivative of the function $\tau(s)$ should be negative in the
method
\begin{eqnarray}
\tau^{\prime}(s)&=&-(\alpha_{2}-2\alpha_{5})-2(\sqrt{\alpha_{9}}+\alpha_{3}\sqrt{\alpha_{8}}\,)
\nonumber \\
&=&-2\alpha_{3}-2(\sqrt{\alpha_{9}}+\alpha_{3}\sqrt{\alpha_{8}}\,)\quad<0.
\end{eqnarray}
From Eqs. (7), (8), (19), and (20), and equating Eq. (7) with the
condition that $\lambda$ should satisfy given by Eq. (9), we find
the eigenvalue equation
\begin{eqnarray}
\alpha_{2}n-(2n+1)\alpha_{5}&+&(2n+1)(\sqrt{\alpha_{9}}+\alpha_{3}\sqrt{\alpha_{8}}\,)+n(n-1)\alpha_{3}\nonumber\\
&+&\alpha_{7}+2\alpha_{3}\alpha_{8}+2\sqrt{\alpha_{8}\alpha_{9}}=0.
\end{eqnarray}
We obtain from Eq. (11) the polynomial $\rho(s)$ as
$\rho(s)=s^{\alpha_{10}-1}(1-\alpha_{3}s)^{\frac{\alpha_{11}}{\alpha_{3}}-\alpha_{10}-1}$
and substituting it into Eq. (10) gives
\begin{eqnarray}
y_{n}(s)=P_{n}^{(\alpha_{10}-1,\frac{\alpha_{11}}{\alpha_{3}}-\alpha_{10}-1)}(1-2\alpha_{3}s)\,,
\end{eqnarray}
where $\alpha_{10}=\alpha_{1}+2\alpha_{4}+2\sqrt{\alpha_{8}}$,
$\alpha_{11}=\alpha_{2}-2\alpha_{5}+2(\sqrt{\alpha_{9}}+\alpha_{3}\sqrt{\alpha_{8}})$
and $P_{n}^{(\alpha,\beta)}(1-2\alpha_{3}s)$ are the Jacobi
polynomials. From Eq. (12), one obtaines
\begin{eqnarray}
\phi(s)=s^{\alpha_{12}}(1-\alpha_{3}s)^{-\alpha_{12}-\frac{\alpha_{13}}{\alpha_{3}}}\,,
\end{eqnarray}
then the general solution $\Psi(s)=\phi(s)y(s)$ becomes
\begin{eqnarray}
\Psi(s)=s^{\alpha_{12}}(1-\alpha_{3}s)^{-\alpha_{12}-\frac{\alpha_{13}}{\alpha_{3}}}
P_{n}^{(\alpha_{10}-1,\frac{\alpha_{11}}{\alpha_{3}}-\alpha_{10}-1)}(1-2\alpha_{3}s).
\end{eqnarray}
where $\alpha_{12}=\alpha_{4}+\sqrt{\alpha_{8}}$ and
$\alpha_{13}=\alpha_{5}-(\sqrt{\alpha_{9}}+\alpha_{3}\sqrt{\alpha_{8}}\,)$.
Let us study the case where the parameter $\alpha_3=0$. In this
type of problems, the eigenfunctions become
\begin{eqnarray}
\Psi(s)=s^{\alpha_{12}}\,e^{\alpha_{13}s}\,L^{\alpha_{10}-1}_{n}(\alpha_{11}s)\,,
\end{eqnarray}
when the limits $lim_{\alpha_3 \rightarrow
0}\,P^{(\alpha_{10}-1\,,\frac{\alpha_{11}}{\alpha_{3}}-\alpha_{10}-1)}_{n}(1-\alpha_{3}s)=
L^{\alpha_{10}-1}_{n}(\alpha_{11}s)$ and $lim_{\alpha_3
\rightarrow
0}\,(1-\alpha_{3}s)^{-\,\alpha_{12}-\frac{\alpha_{13}}{\alpha_{3}}}=
e^{\alpha_{13}s}$ are satisfied and the corresponding energy
spectrum is
\begin{eqnarray}
\alpha_{2}n-2\alpha_{5}n+(2n+1)(\sqrt{\alpha_{9}\,}&-&\alpha_{3}\sqrt{\alpha_{8}\,}\,)+n(n-1)\alpha_{3}
+\alpha_{7}\nonumber\\&+&2\alpha_{3}\alpha_{8}-2\sqrt{\alpha_{8}\alpha_{9}\,}+\alpha_{5}=0\,.
\end{eqnarray}

The generalized Morse potential is given by [44]
\begin{eqnarray}
V_M(x)=De^{-2\beta x}-2De^{-\beta x}\,,
\end{eqnarray}
where $x=(r/r_0)-1$\,,\,$\beta=\alpha r_0$\,,\,$D$ is the
dissociation energy, $r_0$ is the equilibrium distance, and
$\alpha$ is the potential width. The term proportional to $1/r^2$
in Eq. (4) can be expanded about $x=0$ [45]
\begin{eqnarray}
V_M(x)=\,\frac{\kappa(\kappa-1)}{r^2}=\,\frac{a_{0}}{(1+x)^2}=a_{0}(1-2x+3x^2+\ldots)\,;\,\,
a_{0}=\,\frac{\kappa(\kappa-1)}{r_0^2}\,,
\end{eqnarray}
Instead, we now replace $V_M(x)$ by the potential [45]
\begin{eqnarray}
\tilde{V}_M(x)=a_{0}(a_{1}+a_{2}e^{-\beta x}+a_{3}e^{-2\beta
x})\,,
\end{eqnarray}
Expanding the potential $\tilde{V}_M(x)$ around $x=0$, and
combining equal powers with Eq. (28), one can find the arbitrary
constants in the new form of the potential as
\begin{eqnarray}
a_{1}=1-\,\frac{3}{\beta}\,+\,\frac{3}{\beta^2}\,\,;\,\,\,a_{2}=\,\frac{4}{\beta}\,-\,\frac{6}{\beta^2}\,\,;\,\,\,
a_{3}=-\,\frac{1}{\beta}\,+\,\frac{3}{\beta^2}\,.
\end{eqnarray}
Eq. (4) can not be solved analytically because of the last term in
the equation, we prefer to use a mathematical identity such as
$dm(r)/dr=-d\Sigma(r)/dr$ to eliminate this term. We obtain the
mass function from the identity as
\begin{eqnarray}
m(x)=m_{0}+m_{1}e^{-\beta x}+m_{2}e^{-2\beta x}\,,
\end{eqnarray}
where $m_{0}$ corresponds to the integral constant, and the
parameters $m_{1}$, and $m_{2}$ are $2D$, and $-D$, respectively.
The parameter $m_{0}$ will denote the rest mass of the Dirac
particle. By using the potential form given by Eq. (29) replaced
by Eq. (28), inserting the mass function in Eq. (31), setting the
"difference" potential $\Delta(r)$ to generalized Morse potential
in Eq. (27) and using the new variable $s=e^{-\beta x}$,  we have
\begin{eqnarray}
\Big\{\,\frac{d^2}{ds^2}\,+\,\frac{1}{s}\,\frac{d}{ds}\,&+&\frac{1}{s^2}\Big[
-\delta^2(a_{0}a_{1}+m^2_{0}-E^2)-\delta^2[a_{0}a_{2}+(m_{0}-E)(m_{1}+2D)]
s\nonumber\\&-&\delta^2[a_{0}a_{3}+(m_{0}-E)(m_{2}-D]s^2\Big]\Big\}\phi_{n\kappa}(s)=0\,.
\end{eqnarray}
Comparing Eq. (32) with Eq. (13) gives the parameter set
\begin{eqnarray}
\begin{array}{ll}
\alpha_1=1\,, & -\xi_1=-\delta^2[a_{0}a_{3}+(m_{0}-E)(m_{2}-D] \\
\alpha_2=0\,, &
\xi_2=-\delta^2[a_{0}a_{2}+(m_{0}-E)(m_{1}+2D)] \\
\alpha_3=0\,, & -\xi_3=-\delta^2(a_{0}a_{1}+m^2_{0}-E^2) \\
\alpha_4=0\,, & \alpha_5=0
\\ \alpha_6=\xi_1\,, & \alpha_7=-\xi_2 \\
\alpha_8=\xi_3\,, & \alpha_9=\xi_1 \\
\alpha_{10}=1+2\sqrt{\xi_3}\,, & \alpha_{11}=2\sqrt{\xi_1} \\
\alpha_{12}=\sqrt{\xi_3}\,, & \alpha_{13}=-\sqrt{\xi_1}
\end{array}
\end{eqnarray}
where $\delta=1/\alpha$. We write the energy eigenvalue equation
of the generalized Morse potential by using Eq. (26)
\begin{eqnarray}
2\delta\sqrt{a_{0}a_{1}+m^2_{0}-E^2\,}
-\delta\,\frac{a_{0}a_{2}+(m_{0}-E)(m_{1}+2D)}{\sqrt{a_{0}a_{3}+(m_{0}-E)(m_{2}-D)\,}}
=2n+1\,.
\end{eqnarray}
Since the negative energy eigenstates exist in the case of the
pseudospin symmetry [14, 15, 16], so we choose the negative energy
solutions in Eq. (46). In Table I, we give some numerical values
of the negative bound state energies obtained from Eq. (46) for
$CO$ molecule in atomic units, where we use the input parameter
set as $D=11.2256$ eV, $r_{0}=1.1283$ $\AA$, $m_{0}=6.8606719$
amu, and $a=2.59441$ [46], and summarize our results for different
$\tilde{\ell}$, and $n$ values. The corresponding lower spinor
component can be written by using Eq. (25)
\begin{eqnarray}
\phi(s)=s^{w_{1}}\,e^{-\,w_{2}s}L^{2w_{1}}_{n}(2w_{2}s)\,,
\end{eqnarray}
where $w_{1}=\delta\sqrt{a_{0}a_{1}+m^2_{0}-E^2\,}$, and
$w_{2}=\delta\sqrt{a_{0}a_{3}+(m_{0}-E)(m_{2}-D)\,}$.

Let us study the two special limits, pseudospin and spin symmetry
cases, respectively, in the case of the constant mass.
\subsubsection{Pseudospin Case}
The Dirac equation has the exact pseudospin symmetry if the "sum"
potential could satisfy the condition that $d\Sigma(r)/dr=0$, i.e.
$\Sigma(r)=A (const.)$ [14]. The parameters in our formalism
become $m_{1}=m_{2}=0$. Setting the "difference" potential
$\Delta(r)$ to the generalized Morse potential in Eq. (27), using
Eq. (29) for the term proportional to $1/r^2$, and using the new
variable $s=e^{-\beta x}$, we have from Eq. (4)
\begin{eqnarray}
\Big\{\,\frac{d^2}{ds^2}\,+\,\frac{1}{s}\,\frac{d}{ds}\,&+&\frac{1}{s^2}\Big[
-\delta^2[a_{0}a_{1}+M(m_{0}+E)]-\delta^2(2MD+a_{0}a_{2})
s\nonumber\\&+&\delta^2(MD-a_{0}a_{3})s^2\Big]\Big\}\phi(s)=0\,.
\end{eqnarray}
where $M=m_{0}+A-E$. By following the same procedure, the energy
eigenvalue equation for the exact pseudospin symmetry in the case
of constant mass is written
\begin{eqnarray}
2\sqrt{a_{0}a_{1}+M(m_{0}+E)\,}=\frac{a_{0}a_{2}+2DM}{\sqrt{a_{0}a_{3}-DM\,}}+\alpha(2n+1)\,.
\end{eqnarray}
and the corresponding wave functions read as
\begin{eqnarray}
\phi^{m_{1}=m_{2}=0}(s)=s^{w'_{1}}\,e^{-\,w'_{2}s}L^{2w'_{1}}_{n}(2w'_{2}s)\,,
\end{eqnarray}
where $w\,'_{1}=\delta\sqrt{a_{0}a_{1}+M(m_{0}+E)\,}$\,, and
$w\,'_{2}=\delta\sqrt{a_{0}a_{3}-DM\,}$\,. We must consideration
the negative bound states solutions in Eq. (37) because there
exist only the negative eigenvalues in the exact pseudospin
symmetry [14, 15, 16].
\subsubsection{Spin Case}
The spin symmetry appears in the Dirac equation if the condition
is satisfied that $\Delta(r)=V_{v}(r)-V_{s}(r)=A(const.)$. In this
case, we have from Eq. (3)
\begin{eqnarray}
\Big\{\frac{d^2}{dr^2}-\frac{\kappa(\kappa+1)}{r^2}-(m_{0}+E-A)(m_{0}-E-\Sigma(r))\Big\}
\chi(r)=0\,,
\end{eqnarray}
where we set the "sum" potential as generalized Morse potential
given in Eq. (27), and use approximation for the term proportional
to $1/r^2$ in Eq. (29) [45]
\begin{eqnarray}
\tilde{V}_M(x)=b_{0}(b_{1}+b_{2}e^{-\beta x}+b_{3}e^{-2\beta
x})\,,
\end{eqnarray}
where $b_{0}=\kappa(\kappa+1)/r^2_{0}$\,, and the parameters
$b_{i} (i=1, 2, 3)$ are given in Eq. (30). Using the variable
$s=e^{-\beta x}$, and inserting Eq. (40) into Eq. (39), we obtain
\begin{eqnarray}
\Big\{\,\frac{d^2}{ds^2}\,+\,\frac{1}{s}\,\frac{d}{ds}\,&+&\frac{1}{s^2}\Big[
-\delta^2[b_{0}b_{1}+M'(m_{0}-E)]+\delta^2(2DM\,'-b_{0}b_{2})
s\nonumber\\&-&\delta^2(b_{0}b_{3}+DM\,')s^2\Big]\Big\}\chi(s)=0\,.
\end{eqnarray}
where $M\,'=m_{0}+E-A$. We write the energy eigenvalue equation,
and corresponding wave equations in the spin symmetry limit,
respectively,
\begin{eqnarray}
\frac{\delta[2DM\,'-b_{0}b_{2}]}{\sqrt{b_{0}b_{3}+DM\,'\,}}
+2\delta\sqrt{b_{0}b_{1}+M\,'(m_{0}-E)\,}=2n+1\,,
\end{eqnarray}
and
\begin{eqnarray}
\chi^{m_{1}=m_{2}=0}(s)=s^{w''_{1}}\,e^{-\,w''_{2}s}L^{2w''_{1}}_{n}(2w''_{2}s)\,,
\end{eqnarray}
where $w\,''_{1}=\delta\sqrt{b_{0}b_{1}+M'(m_{0}-E)\,}$\,, and
$w\,''_{2}=\delta\sqrt{b_{0}b_{3}+DM'\,}$\,. We must take into
account the positive energy solutions in Eq. (42) in the case of
the exact spin symmetry [14, 15, 16].

In Summary, we have approximately solved the effective mass Dirac
equation for the generalized Morse potential for arbitrary
spin-orbit quantum number $\kappa$ in the position-dependent mass
background. We have found the eigenvalue equation, and
corresponding two-component spinors in terms of Legendre
polynomials by using the parametric NU-method within the framework
of an approximation to the term proportional to $1/r^2$\,. We have
also obtained the energy eigenvalue equations, and corresponding
wave functions for exact pseudospin, and spin symmetry limits in
the case of constant mass. We have observed that our analytical
results in the case of the pseudospin symmetry are good agreement
with the ones obtained in the literature.

\newpage

\begin{table}
\begin{ruledtabular}
\caption{Energy eigenvalues for the $CO$ molecule for different
values of $\tilde{\ell}$ and $(n,\kappa)$ in the case of position
dependent mass.}
\begin{tabular}{ccccc}
$\tilde{\ell}$ & $n$ & $\kappa$ & state & $E<0$ \\ \hline
1 & 1 & -1 & $1s_{1/2}$ & 6.15913020  \\
2 & 1 & -2 & $1p_{3/2}$ & 6.52968379  \\
3 & 1 & -3 & $1d_{5/2}$ & 6.89146288  \\
4 & 1 & -4 & $1f_{7/2}$ & 7.24974882  \\
\end{tabular}
\end{ruledtabular}
\end{table}

\end{document}